\documentclass[12pt,a4paper]{article}

\usepackage[utf8]{inputenc}
\usepackage[T1]{fontenc}
\usepackage{lmodern}
\usepackage{geometry}
\usepackage{times}      
\usepackage{setspace}
\usepackage{amsmath, amsfonts, amssymb, bm, dsfont, relsize, exscale, bigints}

\usepackage{graphicx}
\usepackage{booktabs}
\usepackage{multirow}
\usepackage{ulem} 
\usepackage{tabularx}
\usepackage{array}
\usepackage{makecell}
\usepackage{arydshln}
\usepackage{dcolumn}
\usepackage{float} 
\usepackage{caption}
\usepackage{xcolor}
\usepackage[colorlinks=true,linkcolor=blue,citecolor=blue,urlcolor=teal]{hyperref}

\usepackage{fancyhdr}
\usepackage{url}
\usepackage{appendix}
\usepackage{chngcntr}

\usepackage{caption}
\usepackage{natbib}
\usepackage{xcolor}
\usepackage{hyperref}
\usepackage[all]{hypcap}
\hypersetup{
  colorlinks = true,
  linkcolor = blue,
  citecolor = blue,
  urlcolor  = teal
}

\usepackage{fancyhdr}
\usepackage{url}
\usepackage{pdfpages}
\usepackage{fancybox}
\usepackage{appendix}
\usepackage{chngcntr}

\newcommand{\HRule}{\rule{\linewidth}{0.5mm}}

\begin{document}

\title{
\HRule \\[0.4cm]
{\LARGE \bfseries Global testing of SNP–methylation interactions on binary phenotypes via a logistic functional regression model}\\[0.4cm]
\HRule
}

\author{
Yvelin GANSOU\\
\parbox[t]{0.9\textwidth}{\centering Département de Mathématiques et de Statistique, Université Laval, Québec, QC, Canada}\\
\texttt{yvelin.gansou.1@ulaval.ca}
\and
Karim Oualkacha\\
\parbox[t]{0.9\textwidth}{\centering Département de Mathématiques, Université du Québec à Montréal, Montréal, QC, Canada}\\
\texttt{oualkacha.karim@uqam.ca}
\and
Marzia Angela Cremona\\
\parbox[t]{0.9\textwidth}{\centering Département d’opérations et systèmes de décision, Université Laval, Québec, QC, Canada}\\
\texttt{marzia.cremona@fsa.ulaval.ca}
\and
Lajmi Lakhal-Chaieb\\
\parbox[t]{0.9\textwidth}{\centering Département de Mathématiques et de Statistique, Université Laval, Québec, QC, Canada}\\
\texttt{lajmi.lakhal@mat.ulaval.ca}
}

\date{}
\maketitle


\vspace{1cm}


\begin{abstract} 
\noindent Understanding how genetic and epigenetic factors jointly influence binary health outcomes remains a major challenge in biomedical research. We propose a global test for the overall effect of interactions between DNA methylation and a set of single nucleotide polymorphisms (SNPs) on a binary phenotype. We propose a logistic functional regression model in which methylation measurements at CpG sites are transformed into smooth functional predictors interacting with discrete SNP genotypes through a localized kernel. This framework enables stable inference on region-level interactions while accounting for the spatial structure of methylation around SNPs. Extensive simulations show that the proposed test provides well-calibrated type I error and improved power over classical SNP–CpG pairwise analyses. The practical relevance of the method is illustrated using publicly available methylation and genotyping data from an obesity case–control study.

\end{abstract}

\section{Introduction}
\label{ch3_sec3}

\noindent Binary outcomes such as disease status or case--control indicators are common in genomic and epigenomic studies. Although logistic regression is the standard tool for modeling such outcomes \citep{mccullagh1989generalized}, its direct application to high-dimensional omics data--such as methylation levels measured at thousands of CpG sites together with multiple SNPs--is challenged by the nonlinear logit link, data separation \citep{albert1984existence}, and unstable interaction estimates. These difficulties become even more pronounced when attempting to characterize how genetic and epigenetic factors jointly influence disease susceptibility.

\noindent DNA methylation is an epigenetic modification. It plays a central role in linking environmental exposures to disease \citep{jirtle2007environmental}, regulating tissue differentiation \citep{sandovici2013establishment}, and influencing several complex traits and disorders, including Alzheimer's disease \citep{de2014alzheimer}, obesity-related phenotypes \citep{dick2014dna,voisin2015many}, and diabetes \citep{nilsson2014altered}. DNA methylation is commonly measured using two main technological approaches: array-based platforms, such as the Illumina Infinium BeadChip, which quantify methylation at a fixed set of CpG sites, and sequencing-based technologies, such as targeted or whole-genome bisulfite sequencing, which provide single-base resolution. For an individual $i$, the resulting data consist of methylation proportions $p_{ij}$ observed at genomic positions $t_{ij}$, where both the number and the spacing of CpG sites depend on the technology used. Empirically, methylation levels at neighbouring CpGs tend to exhibit strong spatial correlation. This motivates viewing the discrete observations $\{(t_{ij},p_{ij})\}$ as noisy samples from an underlying smooth methylation function $\Pi_i(t)$ defined along the genomic region. This functional representation of methylation therefore provides a rigorous way to borrow strength across neighbouring CpG sites and reduce the effective dimensionality arising from the large number of measurements within a genomic region \citep{cremona2019functional}. Several studies have demonstrated that capturing regional patterns of methylation carries more biological signal than analyzing individual CpG measurements \citep{hansen2012bsmooth,lakhal2017smoothed,milad2017functional,zhao2020novel}, and that these regional effects may be relevant to disease mechanisms.

\noindent Single nucleotide polymorphisms (SNPs) constitute the most common source of genetic variation in humans. For an individual $i$, SNP data typically consist of genotype counts $G_{id}\in\{0,1,2\}$ recorded at fixed genomic positions $u_d$, where $d$ indexes the SNP and the values 0, 1, and 2 indicate that the individual carries, respectively, zero, one, or two copies of the minor allele. Genome-wide association studies (GWAS) have identified thousands of SNPs associated with cancer, cardiovascular diseases, obesity, and other complex traits \citep{easton2008genome,willer2009six,zeller2012genomewide}. Beyond marginal SNP effects and DNA methylation effects, some recent studies have demonstrated that these two molecular layers may interact in biologically meaningful ways. SNP--methylation interactions have been reported in asthma \citep{soto2013interaction}, triglyceride response to intervention \citep{veenstra2018epigenome}, and newborn telomere length \citep{wang2022genetic}. These findings motivate statistical frameworks capable of capturing interaction effects at the regional level rather than through isolated SNP–CpG pairs \citep{gansou2026functional}.

\noindent A common strategy is to assess SNP--methylation interactions by fitting separate logistic regression models for each SNP--CpG pair and applying multiple-testing corrections such as Bonferroni or FDR (e.g., \citealp{soto2013interaction}).
For simplicity, we refer to this family of procedures as \textit{\textit{LogittestSNPCpG}}. These pairwise strategies ignore the spatial correlation of methylation,
require ad hoc genomic distance thresholds to preselect SNP–CpG pairs, and generate an
overwhelming number of tests, resulting in substantial loss of power. In addition, they are  prone to instability, particularly when certain SNP--CpG combinations perfectly separate cases from controls. These limitations highlight the need
for statistical models that exploit the functional structure of methylation data while
reducing the multiple-testing burden inherent to interaction analyses.

\noindent Functional regression models offer a natural solution, by allowing to treat methylation data as
functional predictor \citep{silverman2005,goldsmith2011penalized,ivanescu2015penalized}.
Prior work has shown that omitting interactions involving functional predictors can lead to
biased inference \citep{usset2016interaction}. However, functional regression methods generally do not accommodate interaction terms between a discrete predictor and a functional
predictor when both lie in the same genomic metric space—that is, when the scalar covariate
(SNP) has its own genomic position and should interact with the functional trajectory
according to their relative distance. To address this limitation, 
\cite{gansou2026functional} filled this methodological gap by introducing a functional linear
model for continuous outcomes in which DNA methylation curves interact with SNP genotypes
through a localized kernel that weights methylation values according to their genomic
distance from each SNP. This approach provides a region-level global test for interaction,
reduces the number of statistical tests, and explicitly incorporates the spatial correlation
among CpG sites.

\noindent In this work, we propose a unified framework for global testing of SNP–methylation interactions via a functional interaction model, which extends the model of \citet{gansou2026functional} to binary phenotypes by developing a logistic functional regression framework in which methylation
trajectories act as smooth functional predictors interacting with SNP genotypes through
distance-based kernel functions. The proposed model is estimated via penalized likelihood within
a generalized linear mixed model (GLMM) formulation, and we derive a likelihood ratio test
to assess the global presence of SNP--methylation interaction effects. This approach
preserves the advantages of functional modeling while addressing the inferential and
computational difficulties specific to logistic regression. This extension to binary outcomes raises several methodological challenges. In particular, logistic models may suffer from separation issues and require appropriate penalization to ensure stable estimation. In addition, inference must account for both the nonlinear link function and the mixed-model structure induced by smoothing. These challenges motivate the methodological developments proposed in this work.

\noindent The remainder of the article is organized as follows. Section~\ref{ch3_sec4} introduces the logistic functional interaction model and the estimation and testing procedures. Simulation results evaluating type~I error and statistical power are presented in Section~\ref{ch3_sec6}. Section~\ref{ch3_sec7} illustrates the method using bisulfite sequencing data from an obesity case–control study. Section~\ref{ch3_sec8} concludes with methodological insights and potential extensions.

\section{Methods}
\label{ch3_sec4}

\subsection{Data and functional representation of methylation data}
\label{ch3_sec2.1}

Let $N$ denote the number of independent individuals for whom DNA methylation has been measured via bisulfite sequencing over a given genomic region. For individual $i \in \{1, \dots, N\}$, let $m_i$ be the number of CpG sites observed, with genomic coordinates $\{t_{ij}\}_{j=1}^{m_i}$. Due to  differences in sequencing coverage, both $m_i$ and the set of genomic positions $\{t_{ij}\}$ may vary across individuals.

\noindent At each CpG site $t_{ij}$, the methylation level is estimated as the ratio $p_{ij}$ of methylated reads to total reads. To capture the spatial structure of methylation along the genome, we convert these methylation levels measured at discrete positions values into a smooth function over the domain of the genomic region.  
For each individual $i$, we represent methylation data as a smooth trajectory $\Pi_i(t)$ obtained via kernel smoothing of the observed CpG measurements. Specifically,
\begin{equation}\label{eq_kernMeth}
\Pi_i(t) = \frac{\sum_{j=1}^{m_i} K\!\left(\frac{t_{ij} - t}{h}\right) p_{ij}}{\sum_{j=1}^{m_i} K\!\left(\frac{t_{ij} - t}{h}\right)},
\end{equation}
where $K$ is a Gaussian kernel. Full details of the smoothing procedure, including the adaptive bandwidth selection, are provided in \citet{gansou2026functional}.

\noindent To simplify computation and interpretation, we rescale the genomic interval to the unit domain $[0,1]$. For each individual, we observe a binary phenotype $Y_i \in {0,1}$, a vector of $D$ SNP genotypes, $G_i = (G_{i1}, \dots, G_{iD})$, where $G_{id} \in {0,1,2}$ for $d = 1, \ldots, D$, with corresponding genomic positions $u_d \in [0,1]$, and a vector of $S$ additional covariates $W_i = (W_{i1}, \dots, W_{iS})$.

\subsection{Logistic functional interaction model}
\label{ch3_sec2.2}
We aim to evaluate whether there exists a joint interaction effect between DNA methylation curves and nearby SNPs on a binary phenotype. Let $\mu_i = \mathbb{P}(Y_i = 1 \mid W_i, G_i, \Pi_i)$ denote the conditional probability that $Y_i = 1$ for individual $i$. Our proposed model takes the following logistic form:
\begin{equation}\label{ch3_eq2}
\text{logit}(\mu_i) = \zeta_0 + \sum_{s=1}^S \zeta_s W_{is} + \sum_{d=1}^D \alpha_d G_{id}
+ \int_0^1 \delta(t)\,\Pi_i(t)\,dt + \sum_{d=1}^D G_{id} \int_0^1 \varphi_d(t,u_d)\,\Pi_i(t)\,dt,
\end{equation}
where $\text{logit}(\mu) = \log\!\left(\frac{\mu}{1-\mu}\right)$, $\zeta_0$ is the intercept, $\{\zeta_s\}_{s=1}^S$ and $\{\alpha_d\}_{d=1}^D$ are coefficients associated with the covariates and SNPs, respectively, $\delta(t)$ measures the main effect of methylation, and $\varphi_d(t,u_d)$ encodes the interaction between the $d^{\text{th}}$ SNP and the methylation profile.\\
To reflect the intuition that interactions are stronger near the SNP position $u_d$, following \citet{gansou2026functional} we define a localized bivariate interaction term:
\begin{eqnarray*}
  \varphi_d (t,u_d) = \eta_d \psi_{\rho} (\vert t - u_d \vert),
\end{eqnarray*}
where $\eta_d$ is a scalar regression coefficient and $\psi_{\rho}: [0,1] \rightarrow [0,1]$ is a decreasing kernel function parameterized by a positive parameter $\rho$, which controls the extent of the interaction that we want to test. 
 For identifiability, we impose $\psi_\rho(0) = 1$. We consider three shapes for $\psi_\rho$, namely:
 convex $\psi_\rho(u) = \exp(-\rho u)$,
 concave $\psi_\rho(u) = \exp(-\rho^2 u^2)$ and
linear $\psi_\rho(u) = \max(1 - \rho u, 0)$.
The resulting model is:
 \begin{eqnarray}\label{ch3_eq3}
   \text{logit}(\mu_i) =\zeta_0 + \sum_{s=1}^S\zeta_s W_{is} + \sum_{d=1}^D \alpha_d G_{id} +    \int_0^1 \delta(t) \Pi_i(t) dt + \sum_{d=1}^D \eta_d G_{id} \int_0^1 \psi_{\rho}( | t - u_d | ) \Pi_i(t) dt.
  \end{eqnarray}

\subsection{Basis expansion and model matrix formulation}
\label{ch3_sec2.3}

To estimate the coefficient function $\delta(t)$, we use a basis expansion based on B-spline functions $\{B_l(t)\}_{l=1}^L$, such that $\delta(t) = \sum_{l=1}^L b_l B_l(t)$, where the coefficients $b_l$ are unknown parameters to be estimated. Our construction follows standard ideas from functional data analysis, in which functional predictors are represented through basis functions and regularized by roughness penalties \citep[see, e.g.,][]{wahba1990spline,silverman2005, reiss2009smoothing,goldsmith2011penalized}. In contrast to the continuous-outcome setting considered in our previous work \cite{gansou2026functional}, the present model embeds the functional slope into a logistic regression framework and relies on a generalized linear mixed model representation for estimation and smoothing-parameter selection \citep{wood2017generalized,crainiceanu2024functional}.\\
Defining $Z_{il} = \int_0^1 B_l(t) \Pi_i(t) dt$, and $\Omega_{id,\rho} = \int_0^1 \psi_\rho(|t - u_d|) \Pi_i(t) dt$, the model can be rewritten as:

\begin{equation}
\text{logit}(\mu_i) = \zeta_0 + W_i^\top \boldsymbol{\zeta} + G_i^\top \boldsymbol{\alpha} + K_{i,\rho}^\top \boldsymbol{\eta} + Z_i^\top \boldsymbol{b},
\end{equation}
where $W_i = (W_{i1}, \ldots, W_{iS})^\top$, $\boldsymbol{\zeta} = (\zeta_1, \ldots, \zeta_S)^\top$, $G_i = (G_{i1}, \ldots, G_{iD})^\top$, $\boldsymbol{\alpha} = (\alpha_1, \ldots, \alpha_D)^\top$, $K_{i,\rho} = (G_{i1}\Omega_{i1,\rho}, \ldots, G_{iD}\Omega_{iD,\rho})^\top$ with $\boldsymbol{\eta} = (\eta_1, \ldots, \eta_D)^\top$, and $Z_i = (Z_{i1}, \ldots, Z_{iL})^\top$ with $\boldsymbol{b} = (b_1, \ldots, b_L)^\top$.
Stacking the observations across individuals yields the generalized linear model
\begin{equation}\label{ch3_eq5}
\text{logit}(\boldsymbol{\mu}) = \boldsymbol{A}_\rho \boldsymbol{\theta},
\end{equation}
where $\boldsymbol{A}_\rho = [\mathds{1} \mid \boldsymbol{W} \mid \boldsymbol{G} \mid \boldsymbol{K}_\rho \mid \boldsymbol{Z}]$, $\boldsymbol{\theta} = (\zeta_0, \boldsymbol{\zeta}^\top, \boldsymbol{\alpha}^\top, \boldsymbol{\eta}^\top, \boldsymbol{b}^\top)^\top$, and $\boldsymbol{\mu} = (\mu_1, \ldots, \mu_N)^\top$. The observed binary outcomes $Y_i$ are assumed independent with
\[
Y_i \sim \text{Bernoulli}(\mu_i), \quad i = 1, \ldots, N.
\]

\subsection{Estimation via penalized likelihood and the PIRLS algorithm for binary outcomes}
\label{ch3_sec5}
To estimate the parameters of the proposed logistic functional regression model, we adopt a penalized likelihood framework for binary outcomes. The smoothness of the functional coefficient $\delta(t)$ is controlled through a 
roughness penalty on its integrated squared second derivative 
\citep{o1986statistical,wahba1990spline,reiss2009smoothing}. 
Such penalties are well known to be equivalent to assuming Gaussian priors on the spline 
coefficients, thereby inducing a mixed–effect representation for $\boldsymbol{b}$ and enabling 
estimation within a generalized linear mixed model (GLMM) framework 
\citep{reiss2009smoothing,wood2017generalized}. Compared to the approach used for continuous outcomes \citep{gansou2026functional}, where a linear mixed model (LMM) is used under normality assumptions, the estimation strategy here addresses the specific challenges of binary data. In particular, the nonlinear link function, the risk of data separation, and the need for stable regularization require a different computational approach based on penalized iterative reweighted least squares (PIRLS) and Laplace approximation \citep{wood2017generalized}.\\
Let $\nu_i = \zeta_0 + W_i^\top \boldsymbol{\zeta} + G_i^\top \boldsymbol{\alpha} + K_{i,\rho}^\top \boldsymbol{\eta} + Z_i^\top \boldsymbol{b}$. The likelihood of the model is given by:
\begin{align*}
L(\boldsymbol{\theta}) 
&= \prod_{i=1}^N \mu_i^{y_i} (1 - \mu_i)^{1 - y_i} \\
&= \prod_{i=1}^{N} \left(\frac{\exp(\nu_i)}{1 + \exp(\nu_i)}\right)^{y_i}
\left(\frac{1}{1 + \exp(\nu_i)}\right)^{1 - y_i}.
\end{align*}
\noindent Hence, the corresponding log-likelihood is
\begin{equation}\label{ch3_eq6}
\ell(\boldsymbol{\theta}) = \sum_{i=1}^N \left[ y_i \nu_i - \log\left(1 + \exp(\nu_i)\right) \right].
\end{equation}
\noindent We consider a roughness penalty applied to the functional coefficient $\delta(t)$, defined as:
\begin{equation}\label{ch3_eq7}
 P(\lambda, \boldsymbol{b}) := \lambda \int_0^1 \left( \frac{\partial^2 \delta(t)}{\partial t^2} \right)^2 dt 
 = \lambda \boldsymbol{b}^\top \boldsymbol{P}_{1p} \boldsymbol{b},
\end{equation}
where $\boldsymbol{P}_{1p}$ is an $L \times L$ positive semi-definite matrix with the $(l, l')$-th element equal to $\int B''_l(t) B''_{l'}(t) dt$. The penalized log-likelihood is then:
\begin{equation}\label{ch3_eq8}
   \ell_p(\boldsymbol{\theta}) = \ell(\boldsymbol{\theta}) -  \frac{1}{2} \boldsymbol{\theta}^\top \boldsymbol{S}_\lambda \boldsymbol{\theta},   
\end{equation}
with $\boldsymbol{S}_\lambda = \mathrm{diag}\{ 0, \boldsymbol{0}_S, \boldsymbol{0}_D, \boldsymbol{0}_D, \lambda \boldsymbol{P}_{1p} \}$.

\noindent For a fixed value of $\rho$, estimation proceeds by treating 
the spline coefficients $\boldsymbol{b}$ as Gaussian random effects, 
while $(\zeta_0, \boldsymbol{\zeta}, \boldsymbol{\alpha}, \boldsymbol{\eta})$ 
are treated as fixed effects. The smoothing parameter $\lambda$ controls 
the smoothness of $\delta(t)$ and is estimated jointly with the model parameters.

\medskip

\noindent We fit the proposed model by expressing it as the following generalized linear mixed model:
\[
\text{logit}(\boldsymbol{\mu}) = \boldsymbol{X} \boldsymbol{\beta} + \boldsymbol{Z} \boldsymbol{b}, 
\quad \boldsymbol{b} \sim \mathcal{N}(\boldsymbol{0}, \boldsymbol{\Psi}_{\lambda}),
\]
where $\boldsymbol{X} = [\mathds{1}_N \mid \boldsymbol{W} \mid \boldsymbol{G} \mid \boldsymbol{K}]$, 
$\boldsymbol{\beta} = (\zeta_0, \boldsymbol{\zeta}^\top, \boldsymbol{\alpha}^\top, \boldsymbol{\eta}^\top)^\top$, 
and $\boldsymbol{\Psi}_{\lambda} = (\lambda \boldsymbol{P}_{1p})^{-1}$.

\medskip

\noindent The marginal likelihood for $\boldsymbol{\beta}$ is given by
\begin{align*}
f(\boldsymbol{y} \mid \boldsymbol{\beta}) 
&= \int f(\boldsymbol{y}, \boldsymbol{b} \mid \boldsymbol{\beta}) \, d\boldsymbol{b} \\
&= \int \exp\!\left\{ \log f(\boldsymbol{y}, \boldsymbol{b} \mid \boldsymbol{\beta}) \right\} d\boldsymbol{b}.
\end{align*}

\noindent A second-order Taylor expansion of $\log f(\boldsymbol{y}, \boldsymbol{b} \mid \boldsymbol{\beta})$ around its maximizer $\hat{\boldsymbol{b}}$ yields
\begin{align*}
f(\boldsymbol{y} \mid \boldsymbol{\beta}) 
&\approx \int \exp \Bigg(
\log f(\boldsymbol{y}, \hat{\boldsymbol{b}} \mid \boldsymbol{\beta})
- \frac{1}{2} (\boldsymbol{b} - \hat{\boldsymbol{b}})^\top 
\boldsymbol{H}_{bb}(\hat{\boldsymbol{b}})
(\boldsymbol{b} - \hat{\boldsymbol{b}})
\Bigg) d\boldsymbol{b},
\end{align*}
where $\boldsymbol{H}_{bb}(\hat{\boldsymbol{b}}) = - \frac{\partial^2 \log f(\boldsymbol{y}, \boldsymbol{b} \mid \boldsymbol{\beta})}{\partial \boldsymbol{b} \partial \boldsymbol{b}^\top} \bigg|_{\boldsymbol{b}=\hat{\boldsymbol{b}}}$ is the negative Hessian matrix.

\medskip

\noindent By Laplace approximation, we obtain
\begin{align*}
f(\boldsymbol{y} \mid \boldsymbol{\beta}) 
&\approx f(\boldsymbol{y}, \hat{\boldsymbol{b}} \mid \boldsymbol{\beta}) 
\,(2\pi)^{L/2}
\left| \boldsymbol{Z}^\top \boldsymbol{\mathbb{W}} \boldsymbol{Z} + \boldsymbol{\Psi}_\lambda^{-1} \right|^{-1/2},
\end{align*}
where
\[
f(\boldsymbol{y}, \hat{\boldsymbol{b}} \mid \boldsymbol{\beta}) 
= f(\boldsymbol{y} \mid \hat{\boldsymbol{b}}, \boldsymbol{\beta}) \,\times f(\hat{\boldsymbol{b}}),
\]
and $\boldsymbol{\mathbb{W}} = \mathrm{diag}\big(\mu_1 (1 - \mu_1), \ldots, \mu_N (1 - \mu_N)\big)$ denotes the working weight matrix arising from the second derivative of the log-likelihood under the logistic model. 

\medskip

\noindent Taking the logarithm, the approximate marginal log-likelihood (for fixed $\lambda$) is
\begin{equation}\label{ch3_eq9}
\ell(\boldsymbol{\beta}) 
\simeq \ell_c(\hat{\boldsymbol{b}}, \boldsymbol{\beta}) 
- \frac{1}{2} \hat{\boldsymbol{b}}^\top \boldsymbol{\Psi}_\lambda^{-1} \hat{\boldsymbol{b}}
- \frac{1}{2} \log |\boldsymbol{\Psi}_\lambda|
- \frac{1}{2} \log \left| \boldsymbol{Z}^\top \boldsymbol{\mathbb{W}} \boldsymbol{Z} + \boldsymbol{\Psi}_\lambda^{-1} \right|,
\end{equation}
where $\ell_c(\boldsymbol{b}, \boldsymbol{\beta}) 
= \log f(\boldsymbol{y} \mid \boldsymbol{b}, \boldsymbol{\beta})$
denotes the conditional log-likelihood.
\medskip

\noindent From equation~(\ref{ch3_eq8}), the estimation of $\boldsymbol{\theta} = (\boldsymbol{\beta}^\top, \boldsymbol{b}^\top)^\top$ has the structure of a generalized linear mixed-model problem. Maximizing the penalized log-likelihood $\ell_p(\boldsymbol{\theta})$ does not yield a closed-form solution, and numerical optimization techniques such as penalized iteratively reweighted least squares (PIRLS) are used to obtain the estimator $\hat{\boldsymbol{\theta}}$.

\noindent The gradient of the penalized log-likelihood evaluated at $\boldsymbol{\theta}^{[k]}$ can be written as
\[
\nabla \ell_p(\boldsymbol{\theta}^{[k]})
=
\boldsymbol{A}^\top \boldsymbol{\mathbb W}\boldsymbol{\mathbb G}(\boldsymbol{y}-\hat{\boldsymbol{\mu}})
-
\boldsymbol{S}_\lambda \boldsymbol{\theta}^{[k]},
\]
and the corresponding penalized Hessian (under the Fisher scoring approximation) is
\[
\boldsymbol{H}_p(\boldsymbol{\theta}^{[k]})
\approx
-
\boldsymbol{A}^\top \boldsymbol{\mathbb W}\boldsymbol{A}
-
\boldsymbol{S}_\lambda,
\]
where $\boldsymbol{\mathbb W}$ is the diagonal weight matrix evaluated at the current fitted mean vector $\hat{\boldsymbol{\mu}}$, which depends on the current iterate $\boldsymbol{\theta}^{[k]}$, and 
\[
\boldsymbol{\mathbb G}
=
\mathrm{diag}\!\left(
\frac{1}{\hat{\mu}_1(1-\hat{\mu}_1)},\dots,
\frac{1}{\hat{\mu}_N(1-\hat{\mu}_N)}
\right)
\]
is the diagonal matrix associated with the derivative of the link function.

\noindent A single Newton--Raphson/Fisher scoring update then takes the form
\[
\boldsymbol{\theta}^{[k+1]}
=
\boldsymbol{\theta}^{[k]}
+
\left(
\boldsymbol{A}^\top \boldsymbol{\mathbb W}\boldsymbol{A}
+
\boldsymbol{S}_\lambda
\right)^{-1}
\left\{
\boldsymbol{A}^\top \boldsymbol{\mathbb W}\boldsymbol{\mathbb G}
(\boldsymbol{y}-\hat{\boldsymbol{\mu}})
-
\boldsymbol{S}_\lambda \boldsymbol{\theta}^{[k]}
\right\}.
\]
Substituting
\[
\boldsymbol{\theta}^{[k]}
=
\left(
\boldsymbol{A}^\top \boldsymbol{\mathbb W}\boldsymbol{A}
+
\boldsymbol{S}_\lambda
\right)^{-1}
\left(
\boldsymbol{A}^\top \boldsymbol{\mathbb W}\boldsymbol{A}
+
\boldsymbol{S}_\lambda
\right)\boldsymbol{\theta}^{[k]}
\]
into the previous expression yields
\[
\boldsymbol{\theta}^{[k+1]}
=
\left(
\boldsymbol{A}^\top \boldsymbol{\mathbb W}\boldsymbol{A}
+
\boldsymbol{S}_\lambda
\right)^{-1}
\boldsymbol{A}^\top \boldsymbol{\mathbb W}
\left\{
\boldsymbol{\mathbb G}(\boldsymbol{y}-\hat{\boldsymbol{\mu}})
+
\boldsymbol{A}\boldsymbol{\theta}^{[k]}
\right\}.
\]
This update can be recognized as the minimizer of the penalized weighted least squares criterion
\[
(\boldsymbol{z}-\boldsymbol{A}\boldsymbol{\theta})^\top
\boldsymbol{\mathbb W}
(\boldsymbol{z}-\boldsymbol{A}\boldsymbol{\theta})
+
\boldsymbol{\theta}^\top \boldsymbol{S}_\lambda \boldsymbol{\theta}
=
(\boldsymbol{z}-\boldsymbol{X}\boldsymbol{\beta}-\boldsymbol{Z}\boldsymbol{b})^\top
\boldsymbol{\mathbb W}
(\boldsymbol{z}-\boldsymbol{X}\boldsymbol{\beta}-\boldsymbol{Z}\boldsymbol{b})
+
\boldsymbol{b}^\top \boldsymbol{\Psi}_\lambda^{-1}\boldsymbol{b},
\]
where
\[
\boldsymbol{z}
=
\boldsymbol{A}\boldsymbol{\theta}^{[k]}
+
\boldsymbol{\mathbb G}(\boldsymbol{y}-\hat{\boldsymbol{\mu}}),
\]
that is, componentwise,
\[
z_i
=
\hat{\nu}_i
+
\frac{y_i-\hat{\mu}_i}
{\hat{\mu}_i(1-\hat{\mu}_i)}.
\]

\noindent Therefore, a penalized iterative reweighted least squares (PIRLS) algorithm is used to compute $\hat{\boldsymbol{\theta}}$.

 \noindent \textbf{Penalized iterative reweighted least squares (PIRLS) algorithm}

\begin{enumerate}

\item Initialize $\hat{\mu}_i^{[0]} \in (0,1)$, for example
\[
\hat{\mu}_i^{[0]} = \frac{y_i+0.5}{2},
\]
and set
\[
\hat{\nu}_i^{[0]}
=
\log\!\left(\frac{\hat{\mu}_i^{[0]}}{1-\hat{\mu}_i^{[0]}}\right).
\]
This choice ensures that $\hat{\nu}_i^{[0]}$ is finite.

\item For $k=0,1,2,\ldots$, repeat the following steps until convergence:

\begin{enumerate}

\item Compute the pseudo-data
\[
z_i^{[k]}
=
\hat{\nu}_i^{[k]}
+
\frac{y_i-\hat{\mu}_i^{[k]}}
{\hat{\mu}_i^{[k]}(1-\hat{\mu}_i^{[k]})},
\]
and the working weights
\[
w_i^{[k]}
=
\hat{\mu}_i^{[k]}(1-\hat{\mu}_i^{[k]}).
\]

\item Update $(\hat{\boldsymbol{\beta}}^{[k+1]},\hat{\boldsymbol{b}}^{[k+1]})$ by minimizing
\[
(\boldsymbol{z}^{[k]}-\boldsymbol{X}\boldsymbol{\beta}-\boldsymbol{Z}\boldsymbol{b})^\top
\boldsymbol{\mathbb{W}}^{[k]}
(\boldsymbol{z}^{[k]}-\boldsymbol{X}\boldsymbol{\beta}-\boldsymbol{Z}\boldsymbol{b})
+
\boldsymbol{b}^\top \boldsymbol{\Psi}_\lambda^{-1}\boldsymbol{b},
\]
then update
\[
\hat{\boldsymbol{\nu}}^{[k+1]}
=
\boldsymbol{X}\hat{\boldsymbol{\beta}}^{[k+1]}
+
\boldsymbol{Z}\hat{\boldsymbol{b}}^{[k+1]},
\qquad
\hat{\mu}_i^{[k+1]}
=
\frac{\exp(\hat{\nu}_i^{[k+1]})}{1+\exp(\hat{\nu}_i^{[k+1]})}.
\]

\end{enumerate}

\end{enumerate}

\noindent If we adopt the Bayesian view of smoothing, in which roughness penalties correspond to Gaussian priors on the penalized spline coefficients, the smoothing parameter $\lambda$ can be estimated by maximizing a Laplace-approximated marginal likelihood criterion \citep{wood2017generalized}. For fixed $\lambda$, the PIRLS algorithm computes the penalized likelihood maximizer $\hat{\boldsymbol{\theta}}$, whereas $\lambda$ is updated in an outer iteration by maximizing an empirical-Bayes criterion of REML type.

\noindent In the Gaussian mixed-model setting, this criterion has the same algebraic form as REML. In the  Bernoulli--logit case, however, the required integral cannot be expressed in closed form and is approximated using Laplace's method. This leads to the following REML-type smoothing-parameter criterion \citep{wood2017generalized}:
\[
V_r(\lambda)
\approx
\ell(\hat{\boldsymbol{\theta}})
-
\frac{1}{2}\hat{\boldsymbol{\theta}}^\top \boldsymbol{S}_\lambda \hat{\boldsymbol{\theta}}
-
\frac{1}{2}\log |\boldsymbol{S}_\lambda|_+
-
\frac{1}{2}\log \left|
\boldsymbol{A}^\top \boldsymbol{\mathbb W}\boldsymbol{A}
+
\boldsymbol{S}_\lambda
\right|
+
\frac{M}{2}\log(2\pi),
\]
where $\ell(\hat{\boldsymbol{\theta}})=\log f(\boldsymbol{y}\mid \hat{\boldsymbol{\theta}})$ is the Bernoulli log-likelihood defined in equation~\eqref{ch3_eq6}, $\boldsymbol{\mathbb W}$ is the diagonal matrix of working weights evaluated at convergence of the PIRLS algorithm, $M$ is the dimension of the null space of $\boldsymbol{S}_\lambda$, and $|\boldsymbol{S}_\lambda|_+$ denotes the product of the non-zero eigenvalues of $\boldsymbol{S}_\lambda$.

\noindent The resulting computational strategy is therefore a nested optimization procedure, with an inner PIRLS loop to estimate $\boldsymbol{\theta}$ for fixed $\lambda$, and an outer iteration to update $\lambda$ by maximizing the above Laplace-approximated REML-type criterion \citep{wood2017generalized}.

\subsection{Hypothesis test of interaction}
\label{ch3_sec2.4}

A key objective of this work is to evaluate whether the inclusion of interaction terms between SNPs and DNA methylation profiles significantly improves the model’s ability to explain the binary outcome. This translates into the following hypothesis testing problem:

\[
H_0 : \eta_d = 0 \quad  \forall d \in \{1, \ldots, D\}, \qquad \text{versus} \quad H_1 : \exists\, d \in \{1, \ldots, D\},  \eta_d \neq 0.
\]

\noindent  Under the null hypothesis, the functional interaction component between each SNP and the methylation curve vanishes, and the model reduces to a standard logistic functional regression without interaction. To formally test $H_0$, we adopt a likelihood ratio test (LRT) approach based on
the Laplace-approximated marginal log-likelihood used in the estimation step.
We compare the marginal log-likelihoods of the full model (including interaction
terms) and the reduced model under $H_0$. The test statistic is
\[
\Lambda = -2 \left( \ell^{\text{reduced}} - \ell^{\text{full}} \right),
\]
where $l^{\text{full}}$ and $l^{\text{reduced}}$ denote the Laplace-approximated
marginal log-likelihoods of the full and reduced models, respectively.
Under $H_0$, the statistic
asymptotically follows a chi-squared distribution with $D$ degrees of freedom:
\[
\Lambda \overset{H_0}{\sim} \chi^2_D.
\]
We reject $H_0$ at level $\alpha$ whenever $\Lambda > \chi^2_{D,1-\alpha}$,
providing a global test for the presence of SNP--methylation interaction effects.

\section{Simulations study}
\label{ch3_sec6}

To evaluate the effectiveness of the logistic functional regression model in assessing SNP--methylation interactions, we designed simulation experiments under both the null and alternative hypotheses. In all simulation scenarios, we fixed the number of covariates at $S = 1$ and set the regression coefficients to $\zeta_0 = 0.9$ and $\zeta_1 = 0.5$. A single covariate was generated for each individual, sampled as $W_1 \sim \mathcal{N}(0, 1^2)$. For the genetic component, we considered $D = 5$ single nucleotide polymorphisms (SNPs), each with a predefined minor allele frequency (MAF) $f \in \{0.14, 0.1, 0.15, 0.08, 0.05\}$. For a given SNP $d$ $(d = 1, \dots, 5)$, we simulated two Bernoulli trials $R_{1d}, R_{2d} \sim \text{Bernoulli}(f)$ and computed the genotype as $G_d = R_{1d} + R_{2d}$, which yields a discrete variable in $\{0, 1, 2\}$. The vectors of SNP and interaction coefficients were defined as 
$\boldsymbol{\alpha} = (-0.61, -0.57, -0.53, -0.49, -0.46)\,(1 + \gamma)$ 
and 
$\boldsymbol{\eta} = (0.9, 0.9, 0.9, 0.9, 0.9)\,\gamma$, 
respectively, where $\gamma$ is a modulation parameter used to define alternative scenarios. 
The genomic positions of the SNPs were fixed at $\{0.09, 0.25, 0.4, 0.6, 0.8\}$ within the unit interval.

\noindent DNA methylation data were constructed using MethylC-seq data from 8 B-cell samples located near the BLK gene on chromosome 8, as described in \cite{lakhal2017smoothed}. The analysis focuses on the genomic region $[11190000, 11460000]$ bp, where differentially methylated regions have been identified. The data preprocessing and functional construction (see equation~(\ref{eq_kernMeth})) follow exactly the same procedure as in \cite{gansou2026functional}. In brief, additional samples are generated by replicating each methylation profile and introducing Gaussian perturbations on the logit scale, resulting in a total of $8(n+1)$ samples.
The function $\delta(t)$ used to simulate functional effects was defined as
\[
\delta(t) = 40 \left( \sqrt{1 + \frac{\gamma}{8}} \right) \cos(3\pi t).
\]
The linear predictor $\nu_i$ was defined as:
\[
\nu_i = 0.9 + 0.5 W_{i1} + \sum_{d=1}^{5} \alpha_d G_{id} + \int_0^1 \delta(t) \Pi_i(t) dt + \sum_{d=1}^{5} \eta_d G_{id} \int_0^1 e^{-\rho|t - u_d|} \Pi_i(t) dt,
\]
where $u_d$ denotes the genomic position of SNP $d$ and the integrals were computed using Riemann sum approximations.\\
Binary responses were generated according to a logistic model, where each observation $Y_i$ follows a Bernoulli distribution:
\[
Y_i \sim \text{Bernoulli}(\mu_i), \quad \text{with} \quad \mu_i = \frac{\exp(\nu_i)}{1 + \exp(\nu_i)}.
\]
The sample size was set to $N \in \{400, 800\}$. A cubic B-spline basis with 10 basis functions was used for estimating $\delta(t)$, and the smoothing parameter $\lambda$ was selected using REML.

\noindent All hypothesis tests were conducted at the nominal significance level of $0.05$.

\subsection{Assessment under the null hypothesis}

To simulate data under the null hypothesis (i.e., absence of SNP--methylation interaction), we set $\gamma = 0$, which implies $\eta_d = 0$ for all $d$. Under this setting, the interaction component vanishes, and the quantities $\psi_\rho$ and $\rho$ do not enter the data-generating mechanism. For completeness, different choices of $\psi_\rho$ and several values of $\rho$ ($0.1$, $1$, and $10$) are considered during model fitting to evaluate the robustness of the procedure.

\begin{figure}[H]
  \centering
  \includegraphics[width=0.8\textwidth]{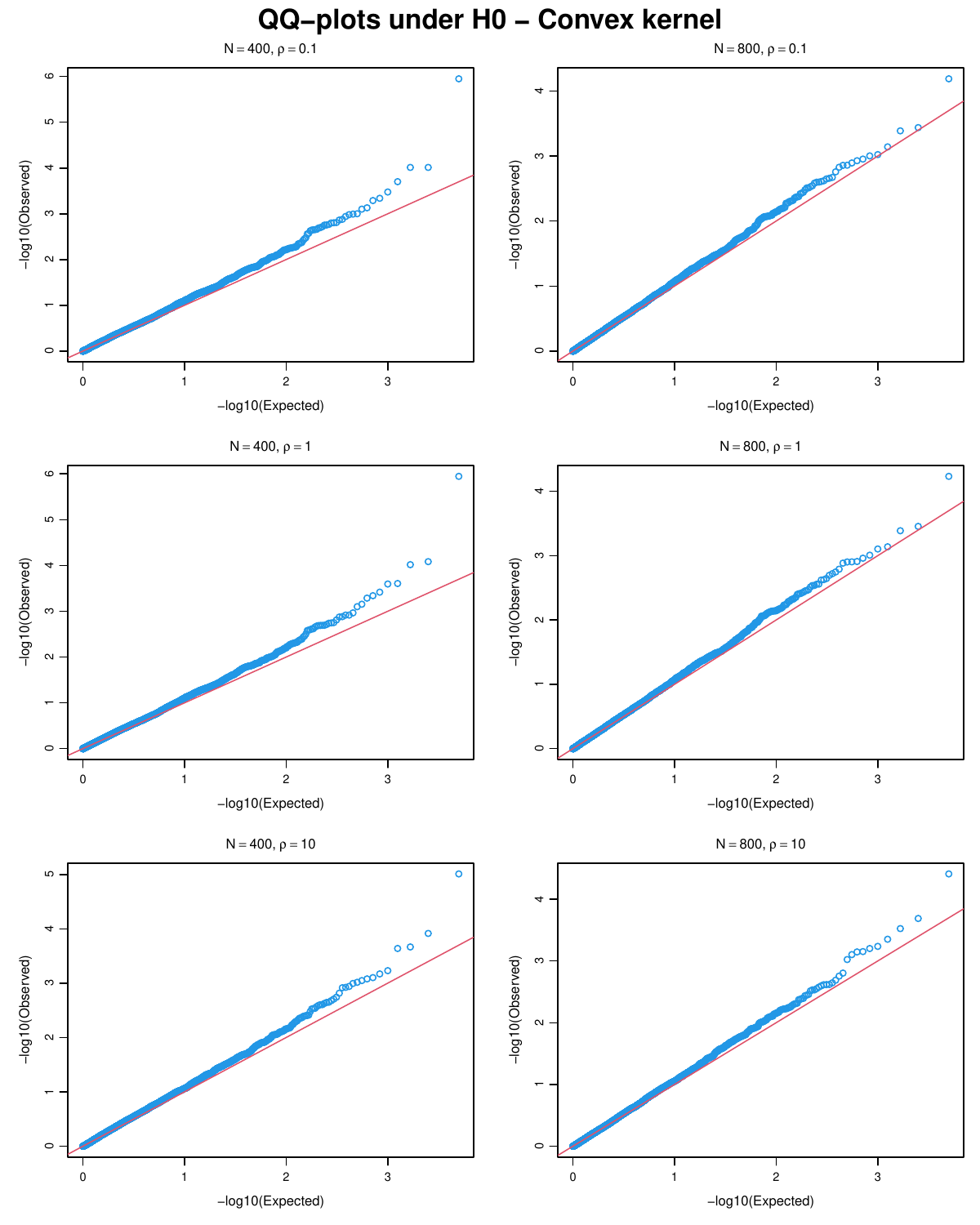}
  \caption{Quantile--quantile plots of p-values from 5000 simulations under the null hypothesis ($H_0$). The model is fitted using the convex kernel $\psi_\rho(u) = e^{-\rho u}$.}
  \label{ch3_fig1}
\end{figure}

\begin{figure}[H]
  \centering
  \includegraphics[width=0.8\textwidth]{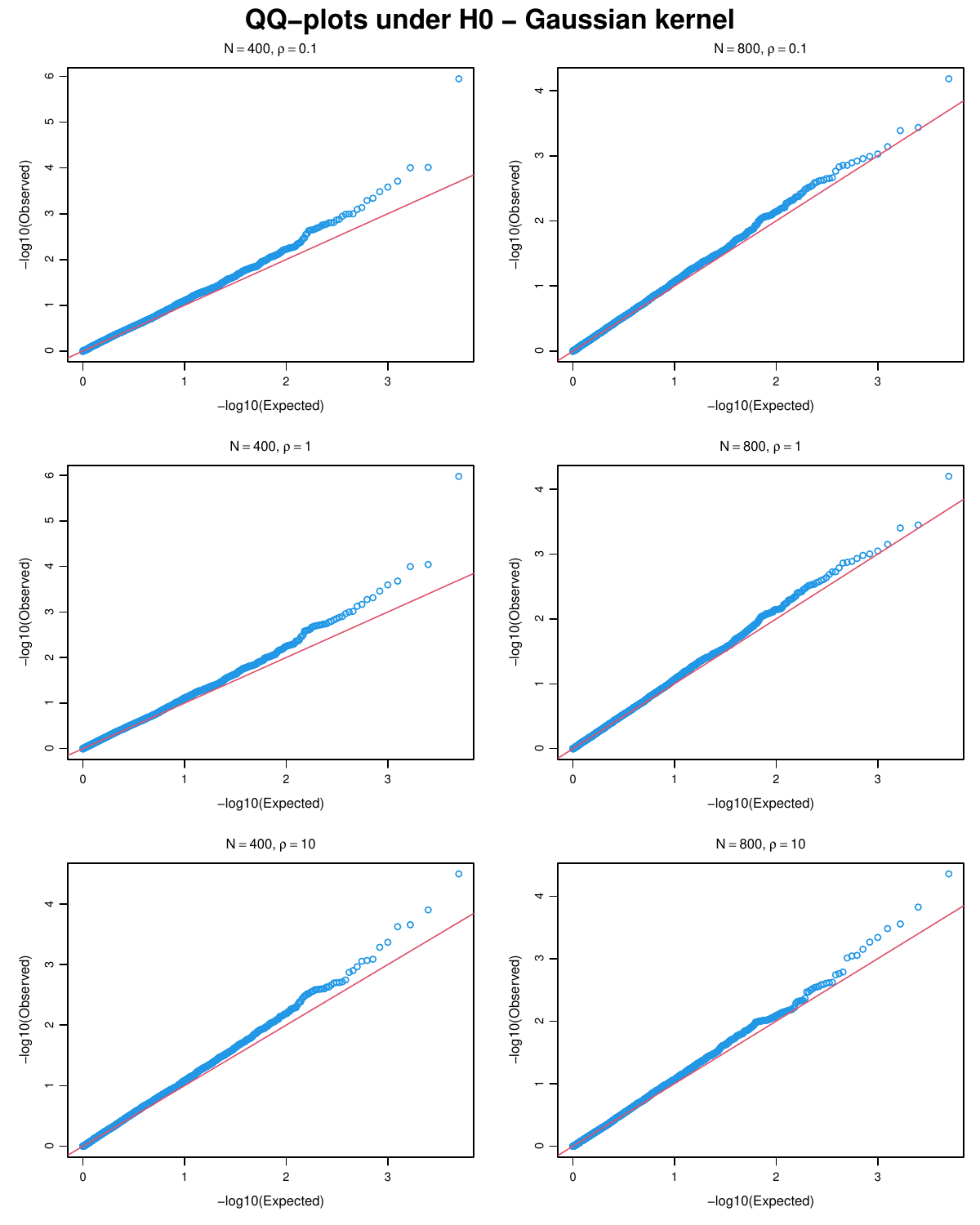}
  \caption{Quantile--quantile plots of p-values from 5000 simulations under the null hypothesis ($H_0$). The model is fitted using the Gaussian kernel $\psi_\rho(u) = \exp(-\rho^2 u^2)$.}
  \label{ch3_fig2}
\end{figure}

\begin{figure}[H]
  \centering
  \includegraphics[width=0.8\textwidth]{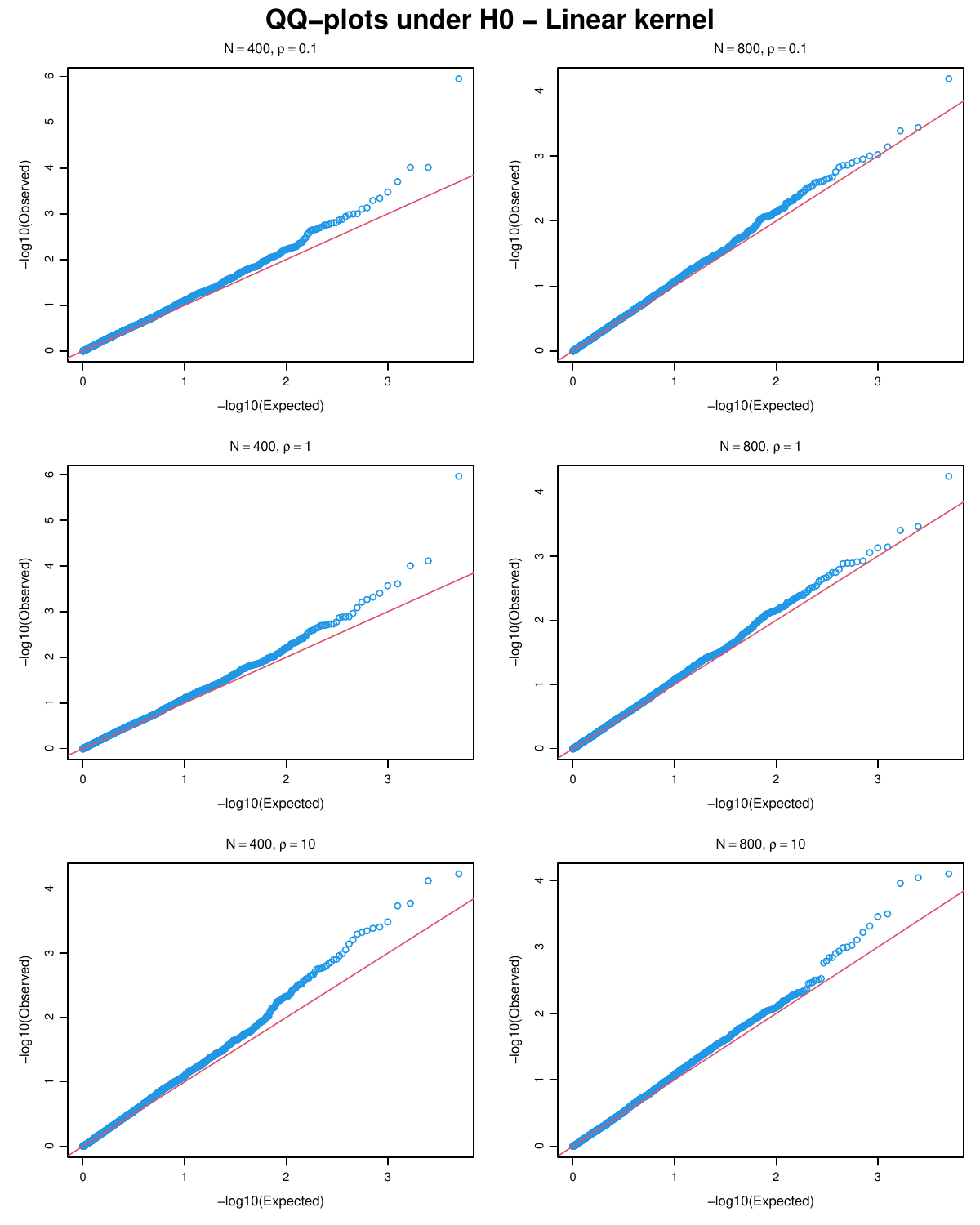}
  \caption{Quantile--quantile plots of p-values from 5000 simulations under the null hypothesis ($H_0$). The model is fitted using the linear kernel $\psi_\rho(u) = \max \{ 1 - \rho u, 0 \}$.}
  \label{ch3_fig3}
\end{figure}

\noindent Figures~\ref{ch3_fig1}--\ref{ch3_fig3} display the quantile--quantile plots of the p-values obtained from $5000$ simulations. These plots indicate that the type I error rate is well controlled by the proposed method across all test settings, especially for larger sample sizes.

\subsection{Evaluation under the alternative hypothesis}

 Simulations under the alternative hypothesis were conducted to assess the power of the proposed test and its sensitivity to misspecification of the parameter $\rho$. Two distinct scenarios were considered.

\begin{figure}[H]   
  \centering  
  \includegraphics[width=0.8\textwidth]{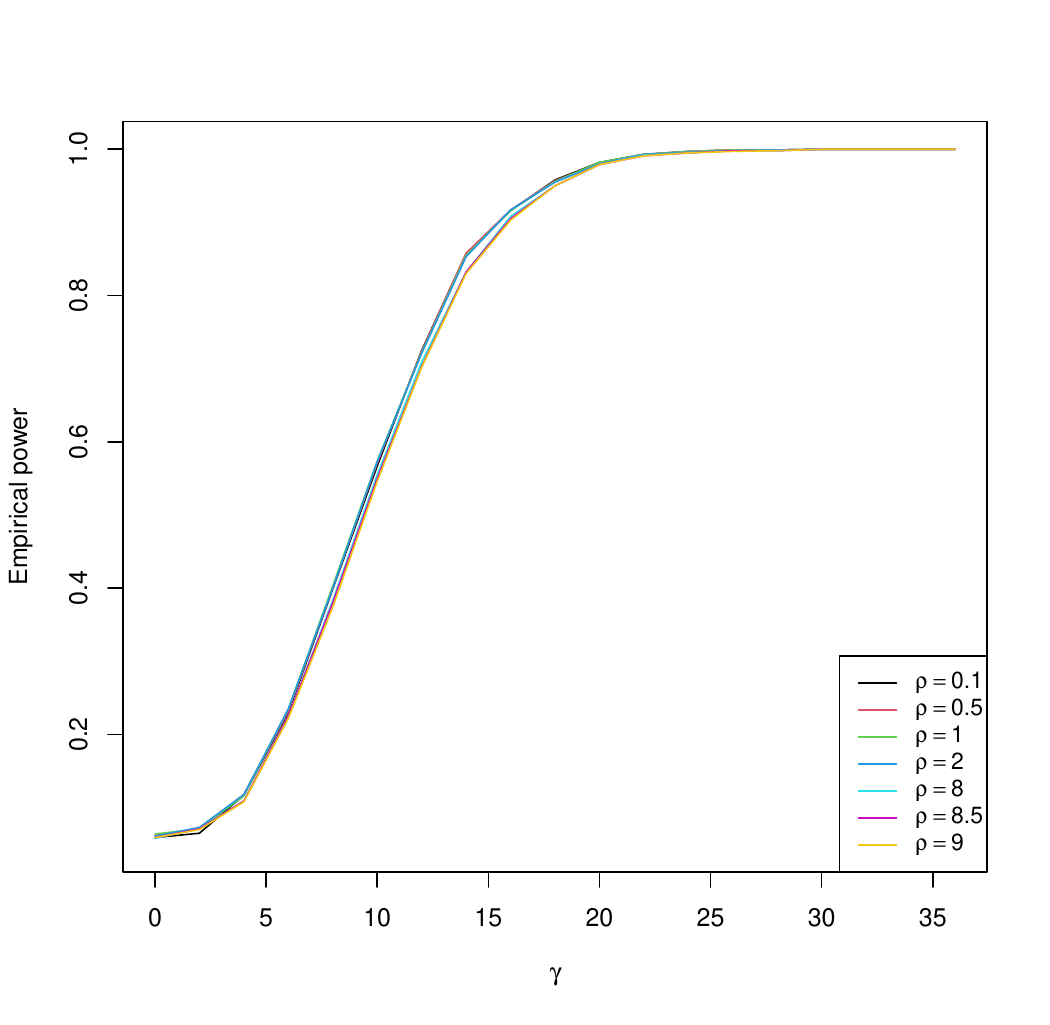}  
  \caption{Empirical power curves as a function of $\gamma$ and $\rho$, with $N = 400$ based on 1,000 simulations. Data were generated under the alternative hypothesis with $\rho = 0.1$, while the values of $\rho$ used for model fitting were misspecified.} 
  \label{ch3_fig4}
\end{figure}

\noindent \textbf{Scenario 1.} Data were generated with $\rho = 0.1$. The parameter $\gamma$ was varied to modulate the strength of the interaction. During testing, the model was fitted using $\psi_\rho(u) = e^{-\rho u}$, with $\rho$ ranging over $\{0.1, 0.5, 1, 2, 8, 8.5, 9\}$. 
Figure~\ref{ch3_fig4} illustrates the empirical power obtained from 1000 simulations. The method exhibits stable performance across the different values of $\rho$, with a maximum power loss of approximately $2.7\%$ when large deviations from the true $\rho$ are introduced during model fitting.

\begin{figure}[H]   
  \centering  
  \includegraphics[width=0.8\textwidth]{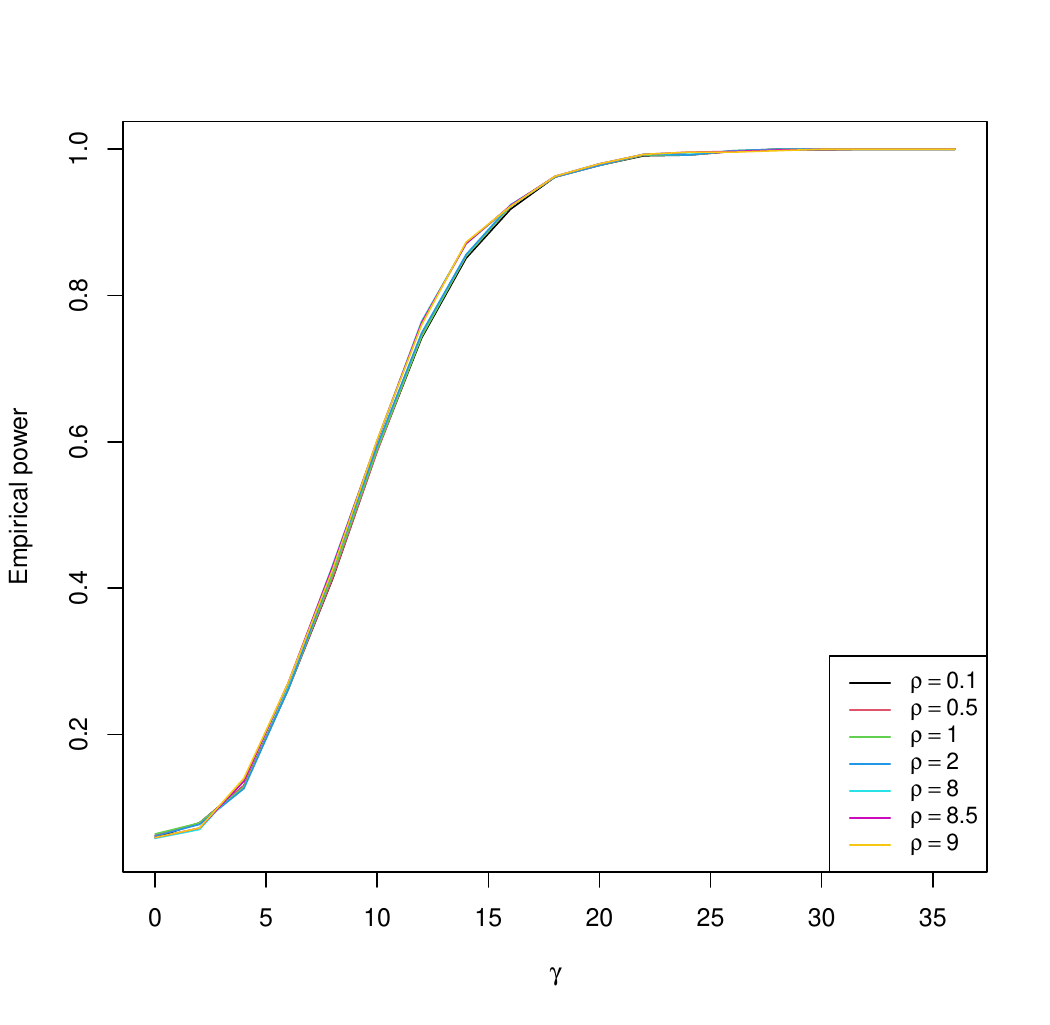}  
  \caption{Empirical power curves as a function of $\gamma$ and $\rho$, with $N = 400$ based on 1,000 simulations. Data were generated under the alternative hypothesis with $\rho = 8$, while the values of $\rho$ used for model fitting were misspecified.} 
  \label{ch3_fig5}
\end{figure}

\noindent \textbf{Scenario 2.} Data were generated with a higher interaction decay parameter, $\rho = 8$. The same set of $\rho$ values was used during model fitting. Results from 1,000 simulations are shown in Figure~\ref{ch3_fig5}. The empirical power again shows limited sensitivity to misspecification of $\rho$, with a maximum observed loss of approximately $2.2\%$.

\section{Methylation, genotype and non genetic covariate data from obesity patients}
\label{ch3_sec7}

To evaluate the practical utility of our proposed logistic functional interaction model, we applied it to the GSE73103 dataset, as described in \cite{voisin2015many}. This dataset comprises DNA methylation and genotyping information collected from a cohort of 355 healthy young Caucasian individuals. Our objective was to assess whether the interaction between obesity-associated SNPs and DNA methylation profiles plays a significant role in the likelihood of being overweight or obese. The original aim of \cite{voisin2015many} was to evaluate whether obesity-associated SNPs identified in large GWAS exert part of their influence through local epigenetic regulation, by testing SNP–CpG associations across chromosomes.

\subsection{Dataset description and preprocessing}

The participants, aged between 14 and 34 years, were divided into two subgroups: 130 individuals aged 14--16 and 225 individuals aged 18--34. Among them, 214 were males and 141 females. Weight status was categorized into three levels: normal-weight, overweight, and obese. Specifically, in the younger subgroup, 101 individuals were of normal weight, 23 overweight, and 6 obese. In the older subgroup, 167 were of normal weight, 47 overweight, and 11 obese. We constructed a binary response variable indicating weight status: normal weight (coded as 0) versus overweight or obese (coded as 1).

\noindent Genome-wide DNA methylation levels were obtained using the Illumina 450K BeadChip, which provides methylation data for 39,869 CpG sites on chromosome 1. However, we identified a genomic gap between positions 121,485,060 bp and 142,618,825 bp, where no methylation measurements were available. Based on this, we partitioned chromosome 1 into two regions: region 1 (positions < 121,485,060 bp) and region 2 (positions > 142,618,825 bp). Region 1 includes five known obesity-associated SNPs (rs984222, rs2815752, rs3934834, rs1514175, rs10783050) and 24,151 CpG sites. Region 2 includes two SNPs (rs516636, rs1011731) and was excluded from our main analysis due to sparse methylation coverage.

 \noindent To obtain smooth representations of the methylation landscape, we transformed the raw methylation beta values into individual-specific functional curves using kernel smoothing, following the approach of \cite{hansen2012bsmooth}. For each individual, the methylation curve $\Pi_i(t)$ was estimated using a Gaussian kernel estimator (see equation~\eqref{eq_kernMeth}), evaluated on a regular grid of 1000 points over the genomic region. An adaptive bandwidth was employed, defined as $h(t)=\max\{d_k(t), h_{\min}\}$, where $d_k(t)$ denotes the distance from $t$ to the $k$th nearest CpG site. We set $k=70$ and $h_{\min}=1$ kb, following the default choices of the \texttt{BSmooth} method \citep{hansen2012bsmooth}. These functional predictors were then used as inputs in our model to capture interactions with discrete SNPs. A descriptive visualization of the average methylation curve across individuals, together with the genomic positions of the SNPs, is provided in \cite{gansou2026functional}, highlighting the spatial structure of the methylation signal along the genomic region.

\noindent Given that cell-type heterogeneity can confound associations between methylation and phenotypes \citep{liu2013epigenome}, we adopted the strategy of \cite{voisin2015many} and adjusted our models using surrogate variables derived from 43 blood-cell-type-discriminative CpG sites. The first two principal components (pca1 and pca2) of these CpGs, which explained over 70\% of the total variance, were included as adjustment covariates.

\subsection{Functional interaction model fitting}

\begin{figure}[H]   
  \centering  
  \includegraphics[width=14cm]{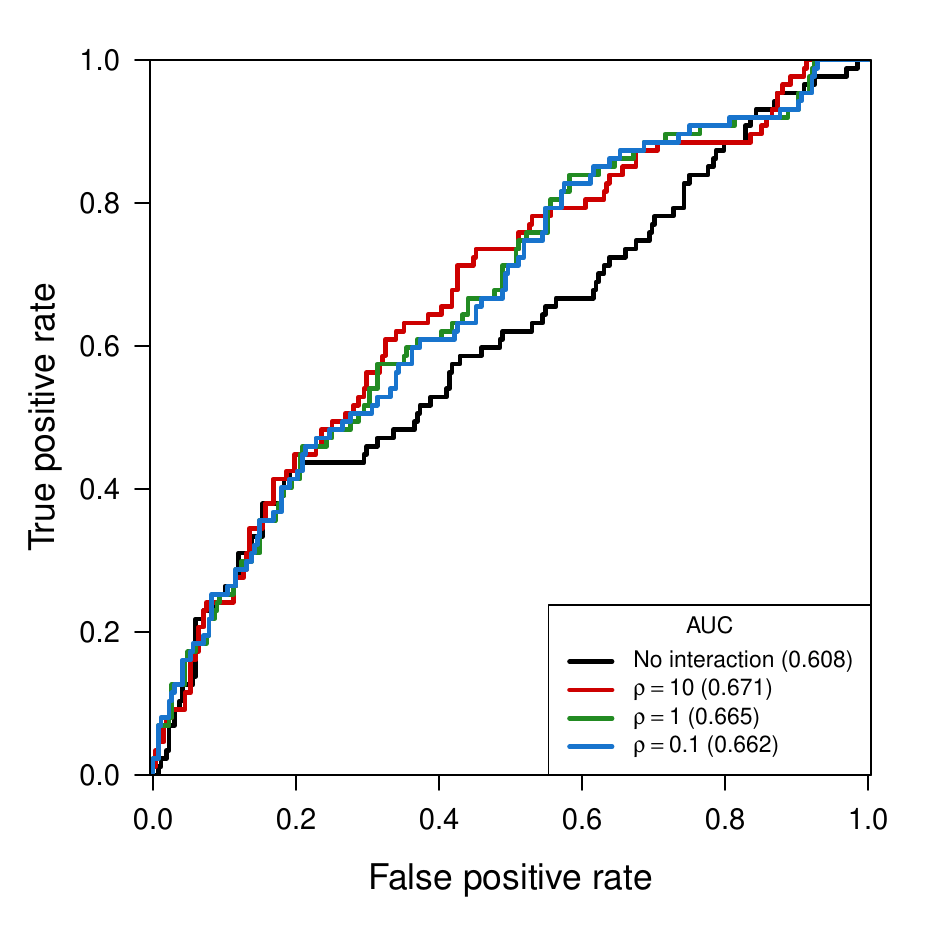} 
  \caption{ROC curves showing improved classification performance for the functional interaction models fitted with different values of $\rho$ ($\rho = 0.1, 1,$ and $10$) compared to the reduced logistic model without SNP–methylation interaction (black curve).} 
  \label{ch3_fig6}
\end{figure}

We modeled the binary weight status as a function of age, sex, pca1, pca2, SNP genotypes, and methylation curves, using our proposed logistic interaction model. To quantify the local interaction between SNPs and methylation, we used the kernel function $\psi_{\rho}(u)=\exp(-\rho u)$ with $\rho=10$, and modeled the interaction effect function $\gamma(t)$ using a cubic B-spline basis with $L = 10$ basis functions.\\
The likelihood ratio test revealed a statistically significant interaction between the five SNPs and methylation curves in region 1, with a $p$-value of 0.04312, indicating a non-negligible contribution of local SNP--methylation interplay to obesity risk.\\
To further investigate the model's predictive performance, we compared it to a reduced logistic functional model that includes the main effects of SNPs and methylation curves but excludes the SNP--methylation interaction term. We fitted our proposed interaction model using three different values of $\rho$ (0.1, 1, and 10), and plotted the receiver operating characteristic (ROC) curves for all models. As illustrated in Figure~\ref{ch3_fig6}, the area under the ROC curve (AUC) was consistently higher for the interaction models, especially for $\rho = 10$, suggesting that incorporating localized functional interactions enhances discrimination between cases and controls.

\subsection{Comparative simulations using real data}

To assess the empirical power of our method in real-data scenarios, we conducted simulations based on chromosome 1 of the GSE73103 dataset. We focused on the SNP rs3934834 and generated synthetic binary response data under three scenarios of increasing interaction complexity:

\begin{itemize}
  \item \textbf{Scenario 1}: Interaction between the SNP and one CpG;
  \item \textbf{Scenario 2}: Interaction between the SNP and 10 CpGs;
  \item \textbf{Scenario 3}: Interaction between the SNP and 20 CpGs.
\end{itemize}
For each scenario, we generated the linear score $\vartheta_i$ from the logistic model:
\[
\vartheta_i = \beta_0 + \beta_{\text{age}} \times \text{Age}_i + \beta_{\text{Sex}} \times \text{Sex}_i + \beta_{\text{SNP}} \times \text{SNP}_i + \beta_{\text{CpG}} \times \text{CpG}_i + \beta_{\text{inter}} \times \text{SNP}_i \times \text{CpG}_i,
\]
and computed the success probability $\text{prob}_i = \frac{e^{\vartheta_i}}{1 + e^{\vartheta_i}}$, followed by simulating $Y_i \sim \text{Bernoulli}(\text{prob}_i)$.\\
To benchmark our approach, we compared it against an existing method referred to as \textit{\textit{LogittestSNPCpG}}, which performs SNP--CpG interaction testing within a 500kb window around the SNP using classical logistic regression and Bonferroni correction for multiple testing over 1080 CpGs.\\
For our method, we fixed $\rho = 10$ and used 10 B-spline basis functions to represent methylation curves. We performed 1,000 replications for each scenario. The true coefficients were defined as follows.

\noindent \textbf{Scenario 1}:
\[
\begin{aligned}
\beta_0 &= 19.129 \sqrt{\gamma_1 \left(1 - \frac{1}{40}\right)}, & \beta_{\text{age}} &= 0.09, & \beta_{\text{Sex}} &= 0.29, \\
\beta_{\text{SNP}} &= -12.64 \sqrt{\gamma_1}, & \beta_{\text{CpG}} &= -36.04 \sqrt{\gamma_1 \left(1 - \frac{6}{100}\right)}, & \beta_{\text{inter}} &= 21.03 \sqrt{\gamma_1}, \\
&\gamma_1 \in \text{seq}(0.1, 2.5, \text{length.out}=10).
\end{aligned}
\]

\begin{figure}[H]   
  \centering  
  \includegraphics[width=1\textwidth]{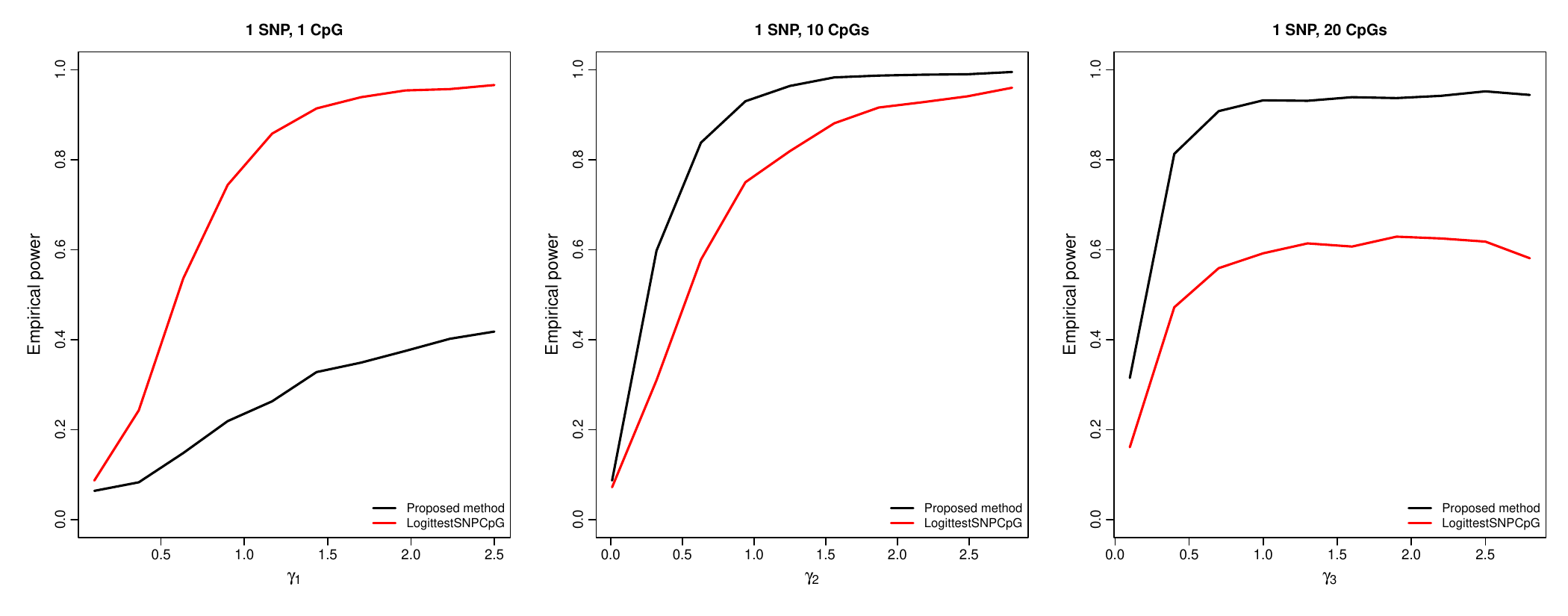}  
  \caption{Empirical power curves as a function of $\gamma_1$, $\gamma_2$, and $\gamma_3$ corresponding respectively to the three scenarios, with sample size $N = 355$ and 1,000 simulations performed. 
  The black curves correspond to the existing method (\textit{LogittestSNPCpG}), while the red curves correspond to the proposed functional interaction method.} 
  \label{ch3_fig7}
\end{figure}

\noindent \textbf{Scenarios 2 and 3} were defined by extending Scenario 1 to multiple CpG sites. In Scenario 2, the linear predictor included 10 CpG main-effect terms and 10 corresponding SNP--CpG interaction terms, whereas Scenario 3 included 20 CpG main-effect terms and 20 corresponding SNP--CpG interaction terms. For scenario $k \in \{2,3\}$, the CpG-related part of the linear predictor was written as
\[
\sum_{j=1}^{m_k} \beta_{\mathrm{CpG}}^{(k)} \mathrm{CpG}_{ij}
+
\sum_{j=1}^{m_k} \beta_{\mathrm{inter}}^{(k)} \mathrm{SNP}_i \times \mathrm{CpG}_{ij},
\]
where $m_2=10$ and $m_3=20$. Within each scenario, the same coefficient was used for all CpG main effects, and the same interaction coefficient was used for all SNP--CpG interaction terms. Both coefficients were scaled by $\sqrt{\gamma_k}$ to progressively increase the strength of the CpG main effects and SNP--CpG interactions, where $\gamma_2$ and $\gamma_3$ each take 10 equally spaced values in $[0.01,2.8]$ and $[0.1,2.8]$, respectively.
The results (Figure~\ref{ch3_fig7}) indicate that our functional interaction model yields higher empirical power compared to \textit{LogittestSNPCpG}, only in scenarios involving multiple CpGs.

\section{Discussion}
\label{ch3_sec8}

This study introduces a novel framework for assessing SNP--methylation interactions in the presence of a binary phenotype, using a logistic functional regression model. Unlike conventional approaches that analyze genomic features separately or rely on simple additive effects, our method captures the joint influence of genotype and methylation through a smooth interaction function modulated by a kernel function.

 \noindent Our simulation results show that the proposed test adequately controls the type I error rate, particularly as the sample size increases, and that its empirical power  increases with the strength of the interaction effects, which is driven by larger values of $\gamma$ and more localized interactions. These findings are consistent with the asymptotic nature of the test. In the real data application, the functional interaction model yielded higher AUC values compared to the corresponding functional model without interaction, indicating that incorporating SNP--methylation interactions provides an additional improvement in predictive performance.

 \noindent It is worth noting that some simulation settings involve relatively strong interaction effects, which may not always reflect realistic effect sizes in biological applications. However, these scenarios were considered within a methodological framework to illustrate the behavior of the proposed test under varying signal strengths. In particular, increases in interaction effects were accompanied by corresponding increases in main effects, allowing us to assess how the empirical power evolves across different configurations.

\noindent A key parameter in the model is $\rho$, which controls the genomic scale over which interactions are captured. Smaller values of $\rho$ correspond to broader interaction windows, assigning relatively higher weights to CpG sites located farther from the SNP, whereas larger values emphasize more localized effects, with higher weights assigned to CpG sites close to the SNP and lower weights to distant ones. Simulation results indicate that moderate misspecification of $\rho$ has a limited impact on performance. In practice, different values of $\rho$ can be considered to assess the sensitivity of the results.

\noindent A limitation of the proposed approach concerns the number of SNPs included simultaneously in the model. Since the response is binary and the model relies on likelihood-based estimation, incorporating a large number of SNPs can lead to instability in parameter estimation due to overparameterization. In particular, when the number of SNPs (and associated interaction terms) becomes large relative to the sample size, standard asymptotic assumptions may no longer hold, and reliable estimation requires the sample size to exceed the effective number of model parameters. Consequently, the proposed method is not intended for genome-wide interaction analysis involving thousands of variants. Rather, the framework is best suited for region-based analyses or for targeted investigations focusing on a moderate number of SNPs, for instance variants previously identified as trait-associated through genome-wide association studies (GWAS). In such settings, the method provides a flexible and biologically interpretable way to assess localized SNP–methylation interactions. Extending the approach to settings with a larger number of SNPs would require additional methodological developments, such as screening procedures or further penalization strategies, which constitute important directions for future research.

\noindent In contrast to linear models previously used for continuous traits \citep{gansou2026functional}, our logistic functional model directly handles binary outcomes without requiring transformation or residual adjustment. This provides a coherent inferential framework and allows the interaction effects to be interpreted in terms of changes in the outcome probability.

\noindent The methodology presented in this paper can be extended in a straightforward way to phenotypes that follow an arbitrary distribution belonging to the exponential family. The inference procedures are then performed within the glmm framework. 

\noindent All methods developped in this work are implemented in the \textrm{R} package \texttt{funInterMethSNP}, available on GitHub at \url{https://github.com/yvelingans/funInterMethSNP}. The package implements models for continuous, binary, and count outcomes, and includes tools to reconstruct methylation curves, fit the models and perform global hypothesis testing.

\noindent Overall, this work contributes to advancing the methodological toolbox for integrative epigenomic analyses, particularly in case–control designs. We anticipate that the approach can be extended to accommodate repeated measurements, or even ordinal outcomes, thereby broadening its applicability in complex trait genetics.

\section{Acknowledgments}
M.A. Cremona is the chairholder of the Chair in Statistical Learning and was funded by the Natural Sciences and Engineering Research Council of Canada (NSERC, grant RGPIN-2020-05657) and the Fonds de recherche du Québec Santé (FRQS, grant 2023-2024-JC-339901).
We thank Aurélie Labbe, Steven Golovkine, and Alexandre Bureau for their comments and suggestions on an earlier version of this work.

 \addcontentsline{toc}{section}{Bibliographie} 

 
\bibliographystyle{apalike}

\bibliography{samplee}

\end{document}